\documentclass[11pt,preprint,preprintnumbers,amssymb]{revtex4}
\usepackage{epsf,epsfig}
\usepackage{subfigure}
\usepackage{dcolumn}
\usepackage{bm}

\evensidemargin=\oddsidemargin

\newcommand{\beq}{\begin{equation}}
\newcommand{\eeq}{\end{equation}}
\newcommand{\bq}{\begin{equation}}
\newcommand{\eq}{\end{equation}}
\newcommand{\ba}{\begin{array}}
\newcommand{\ea}{\end{array}}
\newcommand{\beqa}{\begin{eqnarray}}
\newcommand{\eeqa}{\end{eqnarray}}

\def\End{\end{document}}

\def\to{\rightarrow}

\def\[{\left[}
\def\]{\right]}
\def\({\left(}
\def\){\right)}

\def\l{{\ell}}

\def\U1EM{U(1)_{\rm em}}

\def\[{\left[}
\def\]{\right]}

\preprint{
\hbox to \hsize{
\hfill$\vcenter{\hbox{\bf MAD-PH-09-1551}
         \hbox{December 2009}}$}
}

\begin{document}

\setcounter{footnote}{1}
\renewcommand{\thefootnote}{\fnsymbol{footnote}}




\title{
\vspace*{1cm}
{\Large A Supersymmetric Model with Dirac Neutrino Masses }
}%

\author{\sc {\large Gardner Marshall$^1$}\footnote{grmarshall@wm.edu
},  {\large Mathew McCaskey$^2$}\footnote{
mccaskey@wisc.edu}
, {\large Marc Sher$^{1,2}$}\footnote{mtsher@wm.edu}}
\affiliation{%
$^1$\ Particle Theory Group,~College of William and Mary,
Williamsburg, Virginia 23187\\
$^2$ Department of Physics, 
University of Wisconsin, Madison, Wisconsin 53706
}%

\date{\today}

\begin{abstract}
\vspace*{2mm}
\noindent
New models have recently been proposed in which a second Higgs doublet couples only to the lepton doublets and right-handed neutrinos, yielding Dirac neutrino masses.   The vacuum value of this second ``nu-Higgs" doublet is made very small by means of a very softly-broken $Z_2$ or $U(1)$ symmetry.   The latter is technically natural and avoids fine-tuning and very light scalars.  We consider a supersymmetric version of this model, in which two additional  doublets are added to the MSSM.   If kinematically allowed, the decay of the heavy MSSM scalar into charged nu-Higgs scalars will yield dilepton events which can be separated from the W-pair background.  In addition, the nu-Higgsinos can lead to very dramatic tetralepton, pentalepton and hexalepton events  which have negligible background and can be detected at the LHC and the Tevatron.
\end{abstract}

\maketitle

\baselineskip20pt   

\renewcommand{\baselinestretch}{0.95}
\setcounter{footnote}{0}
\renewcommand{\thefootnote}{\arabic{footnote}}
\setcounter{page}{2}

\addtolength{\topmargin}{-4mm}
\setcounter{page}{2}
\newpage
\section{Introduction}

One of the simplest extensions of the Standard Model, and one of the most studied, is the addition of an additional Higgs doublet \cite{physrep}.    There are many versions of the Two-Higgs Doublet Model (2HDM) in the literature, which differ in the couplings of the Higgs doublets to fermions.   The most familiar are Model I, in which one doublet couples to all fermions and the other couples to no fermions, and Model II, in which one doublet couples to the $Q=2/3$ quarks and the other couples to the $Q=-1/3$ quarks and leptons.  The latter is a feature of supersymmetric models \cite{haberkane}.   These couplings are generally restricted by a $Z_2$ symmetry.   Another model, Model III, has no such discrete symmetry but then also has tree-level flavor-changing neutral currents \cite{model3}.

Another version of the 2HDM has one doublet coupling to all of the quarks and the second doublet coupling to the leptons.   Although the basic structure of this model was proposed long ago \cite{2higgs}, it has received a resurgence of interest \cite{more2higgs} due to the existence of non-zero neutrino masses.   A modified version, in which one doublet couples to all of the quarks and charged leptons, and the second doublet couples only to the left-handed lepton doublet and the right-handed neutrino was proposed by Gabriel and Nandi \cite{gn}.   The model will allow for Dirac neutrino masses, which are small due to a very small (less than an eV) vacuum expectation value for the second doublet (as opposed to the one-doublet case in which they are small due to very small Yukawa couplings).    The model has a $Z_2$ symmetry in which the second doublet and the right-handed neutrinos are odd and all other fields are even.

The model of Gabriel and Nandi \cite{gn} has the feature that it contains a very light scalar which causes problems with standard cosmology.  This feature remains even if the $Z_2$ symmetry is promoted to a $U(1)$ (which can eliminate Majorana mass terms for the right-handed neutrinos).   Very recently, Davidson and Logan (DL) proposed \cite{dl} enforcing the coupling structure with a global $U(1)$, but breaking the $U(1)$ explicitly, through a dimension-2 soft term in the Higgs potential.   This avoids any Goldstone bosons and other light scalars, and only requires that the soft term have a magnitude of approximately an ${\rm MeV}^2$.   Since this term is the only $U(1)$ breaking term, the smallness of its size is technically natural.   The charged Higgs boson in the model has very distinctive signatures, decaying into a left-handed charged lepton and a neutrino, with branching ratios determined by the Pontecorvo, Maki, Nakagawa, Sakata (PMNS) neutrino mixing matrix.

In this paper, we consider the supersymmetric version of the Davidson-Logan model.  We extend the MSSM by adding two additional Higgs doublets of opposite hypercharge, which have opposite quantum numbers under the global $U(1)$.   The right-handed neutrinos are also charged under the $U(1)$ and all other fields are neutral.   We first show that, through D-terms, the heavy neutral scalar of the MSSM will decay (if kinematically allowed) into the lightest of the $U(1)$-charged Higgs, leading to distinctive signatures.   We then look at the supersymmetric partners of the $U(1)$-charged Higgs, and find some remarkable signatures at the LHC and the Tevatron, including dramatic multilepton events which have little to no background and yet accessible cross sections.

\section{The Model}

To ensure anomaly cancellation, we must add an even number of doublets to the MSSM.    Referring to the MSSM doublets as $H_1$ and $H_2$, we add two new doublets, $H_3$ and $H_4$ and three right-handed neutrino superfields, $N_i$.   A global $U(1)$ symmetry is imposed in which $H_3$, $H_4$ and $N_i$ have charges $-1$, $+1$ and $-1$ respectively, while all other fields are uncharged.   As usual, R-parity will be conserved in the model.

The superpotential is 
\begin{equation}
W = W_0 + y^\prime N L H_4 +  \mu_{34} H_3 H_4 
\end{equation}
\noindent where $W_0$ is the MSSM superpotential and $L$ is the left-handed lepton doublet (generation indices are understood).  Since the new Higgs fields only couple to right-handed neutrinos and leptons, we refer to them as ``nu-Higgs."   Due to the global $U(1)$ symmetry, there is no mixing with the MSSM Higgs bosons at this stage.  We also assume that the mass-squared parameters of the nu-Higgs fields are sufficiently large and positive that the fields do not acquire vacuum expectation values (vevs).    DL show that there is a cosmological lower bound of $1/30$ on $y^\prime$; our results will not be sensitive to $y^\prime$ as long as it is not extremely small (so that decays occur at the vertex).

Following Davidson and Logan, the global $U(1)$ is broken explicitly by soft dimension-2 terms in the potential $\mu_{14}^2 H_1 H_4 + \mu_{23}^2 H_2 H_3$.   These terms will result in vevs for the nu-Higgs fields which have the seesaw form $v_3 = \mu_{23}^2 v_2/M_A$ and $v_4 = \mu_{14}^2 v_1/M_A$, where $M_A$ is generally the weak scale.   Since the vevs of $H_3$ and $H_4$ cannot be too dissimilar (due to the $U(1)$-conserving $\mu_{34}$ coupling), we must have $v_3\sim v_4\sim$ eV in order to give the correct neutrino masses without very small Yukawa couplings.  This implies that $\mu_{14}$ and $\mu_{23}$ are both of the order of an MeV.

Since these soft terms are the only source of $U(1)$ breaking, their small size is technically natural.~\cite{froggatt}  One can easily imagine getting such small terms through a Froggatt-Nielsen mechanism.  For example, introducing a flavon field $F$ with a $U(1)$ charge of $1/3$ would allow one to include an effective $U(1)$ invariant operator ${1\over M^3}F^3 \mu^2 H_2 H_3$, where $M$ is the flavon scale and $\mu$ is the weak scale.  Then for $\langle F\rangle/M \sim 0.05$, one obtains an effective term of the correct size.   We will not pursue this issue further, but just note that it may not be too difficult to explain the size of these small terms.

This model does seem rather ad hoc, and one can ask about a possible ultraviolet completion, possibly in the context of a grand unified theory.   Of course, this question can be asked about any leptophilic model, whether supersymmetric or not.    As noted earlier, there are many discussions of leptophilic models~\cite{2higgs,more2higgs} in which the charged leptons (and the left-handed neutrino) have a different sign under the $Z_2$ symmetry than the quarks.   It is difficult to see how to reconcile these models with grand unification, since the charged leptons and quarks are generally in the same representation.    In this model, however, the charged leptons and quarks have the same sign under the $Z_2$, and thus it is easy to incorporate into a GUT.  For example, in SU(5), one could simply promote the $H$ fields to $5$-plets, and the $N$ fields to singlets, and impose the same discrete symmetry.   Whether one can find a model which can also give Froggatt-Neilsen type terms discussed in the previous paragraph is unclear---it would possibly be more contrived than the model we have presented.   Without such a detailed model, discussion of sparticle mass spectra would be premature.  Nonetheless, this model would appear to be easier to embed in a GUT than other leptophilic models.

The nu-Higgs spectrum consists of two scalars, two pseudoscalars, two pairs of charged scalars, two neutral nu-Higgsinos and a pair of charged nu-Higgsinos.   What are their masses?  In the scalar sector, DL had several unknown quartic couplings, and thus the relative masses of the scalars was arbitrary; they had to consider cases in which the charged scalar was either heavier than or lighter than the neutral scalars, which affected the phenomenology.  In this case, however, the quartic terms are completely determined by gauge couplings.  Ignoring the very small corrections due to $U(1)$ breaking, the masses were calculated in Ref. \cite{aranda}.   They found that the neutral scalar mass matrix depended on unknown parameters, but the pseudoscalar mass matrix was identical to the scalar, and the charged scalar mass matrix only differed slightly.   It was shown that the lightest charged scalar was a few GeV heavier than the degenerate scalar and pseudoscalar.    Since the masses are similar, decays of the nu-Higgs particles into each other will be phase-space forbidden or suppressed, and the decays into leptons will predominate.   As a result, we will focus on the lighter of the states, and take the scalar, pseudoscalar and charged scalar to be degenerate in mass (for phenomenological purposes, keeping in mind that the charged scalar is slightly heavier).   They will be referred to as $\chi^0, \chi_A$ and $\chi^\pm$ respectively.  The nu-Higgsino masses depend only on $\mu_{34}$ and are completely degenerate at tree level.   Note that the neutral nu-Higgsino mass matrix is completely off-diagonal, giving a mixing angle of $45$ degrees when rotating to mass eigenstates.   Thus, both of the neutral nu-Higgsinos as well as the charged nu-Higginos are degenerate in mass.   For notational simplicity, we refer to them as $ \widetilde{\chi}^0$ and $ \widetilde{\chi}^\pm$.  Although there are two neutral states, the branching ratios will be the same for each, and we will account for the factors of two in the production cross section.

Focusing on the lighter of the nu-Higgs and on the nu-Higgsinos, we can look at their decays.     The decays of the neutral nu-Higgs, $\chi^0, \chi_A$  are into neutrinos, and are thus unobservable.  The charged nu-Higgs, $\chi^{\pm}$, will decay into all nine combinations $\l_i\nu_j$.  Davidson and Logan show that the decay rate into $l^{\pm}\nu$ is proportional to $\sum_i m_{\nu_i}^2|U_{li}|^2$, where $U$ is the PMNS mixing matrix.    For a normal neutrino mass hierarchy, with the lightest neutrino having a mass below $10^{-3}$ eV, the decays will be into $\mu\nu$ and $\tau\nu$ with branching ratios between $40$ and $60$ percent.   For an inverted hierarchy, the rate into $e\nu$ is about $50$ percent, with the balance shared equally between $\mu\nu$ and $\tau\nu$.  A discussion of the PMNS mixing matrix parameters used is in Appendix~\ref{appendix}.   The actual width is narrow, but the decay still occurs at the vertex.

The nu-Higgsinos decay as
\beqa
 \widetilde{\chi}^0 & \to & \nu_L +  \widetilde{\nu}_R , \  \nu_R +  \widetilde{\nu}_L \cr
 \widetilde{\chi}^{\pm} & \to & l +  \widetilde{\nu}_R, \ \nu_R +  \widetilde{l}
\eeqa
with equal branching ratios in the limit that the slepton and sneutrino have equal mass.  Note that because the  nu-Higgsino mass eigenstates are made up of equal parts of $ \widetilde{H}_3$ and $ \widetilde{H}_4$, and the former does not have any two body decays, both mass eigenstates will decay similarly.

Finally, the right-handed sneutrino, $ \widetilde{\nu}_R$ can only decay through a virtual nu-Higgs (plus a left-handed sneutrino or slepton) or a virtual nu-Higgsino (plus a left-handed neutrino or lepton), and has no two body decays.   As a result, it can have visible decays, and thus the $ \widetilde{\chi}^0$ decays above could be observable through both decay chains.

We now turn to the detailed phenomenology of these nu-Higgs particles at the LHC and the Tevatron.

\section{Phenomenology of the nu-Higgs scalars}

In the non-supersymmetric version of the model, the phenomenology of the nu-Higgs scalars was discussed by Davidson and Logan~\cite{dl}.    As noted above, since supersymmetry did not restrict the masses in their model, they needed to consider a range of possibilities.  In the supersymmetric version, the lightest neutral nu-Higgs is only slightly lighter than the lightest charged nu-Higgs.  As a result, it will only decay into neutrinos and would thus be unobservable\footnote{We are assuming here that its supersymmetric partners are not much lighter--if so, other two-body decay channels which are observable could open up}.     The charged nu-Higgs will decay into a charged lepton and a neutrino; summing over the unobserved neutrino leads to the charged lepton being approximately $50\%~\mu$, $50\%~\tau$ ($50\%~e$, $25\%~\mu$, $25\%~\tau$) for the normal (inverted) hierarchy.

This was discussed by Davidson and Logan, and the primary production mechanism was through a Drell-Yan photon or $Z$.   In the supersymmetric version, a new production mechanism opens up, and can be substantially larger.   Because of  the $SU(2)$ and $U(1)$ D-terms, there is an interaction $(H_1^\dagger H_1-H_2^\dagger H_2)(H_3^\dagger H_3 - H_4^\dagger H_4)$, and when $H_1$ and $H_2$ acquire vev's, this will lead to a three point interaction between an MSSM Higgs and two nu-Higgs scalars.    The lightest MSSM Higgs is too light to decay into two charged nu-Higgs, but the heavier one, $H$, is not.   One can thus produce the $H$ of the MSSM via gluon fusion~\cite{spira}, and it will then decay into a pair of charged nu-Higgs scalars.  Each of those will decay into a charged lepton and a neutrino.

For the production of the heavy MSSM Higgs through gluon fusion we use $\tan\beta=3$.    In order to calculate the pair production of the charged nu-Higgs bosons we first calculate the branching fraction of the heavy MSSM Higgs to a pair of charged nu-Higgs bosons.    From the D-terms in the potential we can derive the coupling of the heavy MSSM Higgs to the charged nu-Higgs.
\begin{equation}
g_{H^{0}\chi^{+}\chi^{-}} = \left(\frac{gM_{Z}}{2\cos\theta_{W}}\right)\cos(2\theta_{W})\cos(2\gamma)\cos(\alpha+\beta)
\end{equation}

\noindent where $\alpha$ and $\beta$ are the standard MSSM parameters and $\gamma$ is the mixing angle that diagonalizes the neutral nu-Higgs mass-squared matrix.    The decay width of the MSSM Higgs to charged nu-Higgs bosons then can be written as
\begin{equation}
\Gamma(H^{0}\to \chi^{+}\chi^{-})=\frac{g_{H^{0}\chi^{+}\chi^{-}}^{2}}{16\pi M_{H^{0}}}\sqrt{1-\frac{4M_{\chi^{+}}^{2}}{M_{H^{0}}^{2}}}.
\end{equation}

For example using an MSSM Higgs of $400$ GeV and an charged nu-Higgs mass of $100$ GeV we calculate the production cross section of a pair of charged nu-higgs bosons to be $14$ fb assuming $\cos(2\gamma)=1$.    Since both of the charged nu-Higgs bosons will each decay into a charged lepton and neutrino, the collider signature will be two leptons (which generally can be different) and missing energy.

There are two backgrounds that we consider.    The primary background is $W$-pair production with both $W$ bosons decaying leptonically.    We also consider $t\bar{t}$ pair production as another possible background.    Due to its large production cross section ($\sim\mathcal{O}$($100$ pb)) it is possible for $t\bar{t}$ events to have both tops decay semi-leptonically and both $b$-quarks be missed by the detector because of too large rapidity or too small transverse momentum (i.e. $\eta_b>2.0$ and $p_{T,\text{b}}<20 $ GeV, respectively).

We generated events for the signal and both of the backgrounds using the MadGraph/MadEvent program package~\cite{madgraph} using the CTEQ6L parton distribution functions~\cite{CTEQ}.    To simulate these events in a collider we apply the following acceptance and isolation cuts on the two final state leptons:
\begin{eqnarray}
p_{T,\text{min}}=20~\text{GeV},\\
|\eta_{e}|<2.4,~|\eta_{\mu}|<2.1,\\
\Delta R=\sqrt{\Delta\eta^{2}+\Delta\phi^{2}}>0.4,
\end{eqnarray}

\noindent where $\eta$ and $\phi$ are the pseudorapidity and azimuthal angle of a final state lepton respectively.  We also apply a $90\%$ tagging efficiency for the leptons (electrons and muons).    For the final state leptons we do not include taus because of the small $40\%$ tau tagging efficiency and a relatively large light jet mis-tagging rate $\sim\mathcal{O}$($1\%$)~\cite{ATLAS-TDR}.   

To extract the signal from our background we first require that there are two final state leptons that pass the acceptance and isolation cuts.    This naturally reduces the number of $t\bar{t}$ events.    Because our final state particles decay from a heavy Higgs boson, we expect that there should be a large amount of missing transverse energy from the two energetic neutrinos as well as a large total transverse mass from the high $p_{T}$ leptons.    We exploit this by requiring the missing transverse energy be greater than $90$ GeV and the total transverse mass be greater than $250$ GeV.    Applying these collider cuts using the MadAnalysis program package~\cite{madgraph} we obtain the results given in Table~\ref{tab:collider}.    With $100$ fb$^{-1}$ of integrated luminosity we obtain a statistical significance near $3\sigma$ for both values of the charged nu-Higgs mass and a heavy MSSM Higgs mass of $400$ GeV.    For an MSSM Higgs of $300$ GeV and charged nu-Higgs mass of $100$ GeV we obtained a statistical significance over $5\sigma$ due to the larger production cross section of the MSSM Higgs and larger branching fraction to a charged nu-Higgs pair.  Note we did not assume any particular hierarchical structure of the neutrino masses.   If we knew more specifically what the flavor distribution of the leptons in the charged nu-Higgs  decay, it may be easier to extract the signal from backgrounds.

\begin{table}[tpb]
\begin{center}
\begin{tabular}{|c|c|c||c|c||c|c||c|c|}
\hline
Cut & $W^{+}W^{-}$ & $t\bar{t}$ & $M_{H}=400$ GeV & $\frac{S}{\sqrt{S+B}}$ & $M_{H}=400$ GeV & $\frac{S}{\sqrt{S+B}}$ & $M_{H}=300$ GeV & $\frac{S}{\sqrt{S+B}}$ \\
 & & & $M_{\chi} = 100$ GeV & & $M_{\chi} = 150$ GeV & & $M_{\chi} = 100$ GeV & \\ 
\hline\hline
tagging: $2\ell$, $0$ jets & $113000$ & $610$ & $420$ & $1.2\pm0.06$ & $300$ & $0.9\pm0.05$ & $2200$ & $6.5\pm0.1$ \\
$E_{T,\text{miss}}>90$ GeV & $1700$ & $160$ & $120$ & $2.7\pm0.2$ & $150$ & $3.3\pm0.3$ & $400$ & $8.4\pm0.4$ \\
$M_{TR}>250$ GeV & $1300$ & $160$ & $110$ & $2.8\pm0.3$ & $130$ & $3.3\pm0.3$ & $210$ & $5.1\pm0.3$ \\
\hline
\end{tabular}
\caption{Collider cuts used to extract the two lepton, two neutrino signal from $\chi^{+}\chi^{-}$ production from the backgrounds considered.    With $100$ fb$^{-1}$ of integrated luminosity we obtain a statistical significance over $5\sigma$ for the $M_{H}=300$ GeV case and near $3\sigma$ for $M_{H}=400$ GeV and both values of $M_\chi$.    It is interesting to note that for the charged nu-Higgs of $150$ GeV that even though the production cross section is smaller, the larger nu-Higgs yields events with larger missing energies and total transverse mass which allows more events to pass our collider cuts.}
\label{tab:collider}
\end{center}
\end{table}

\section{Phenomenology of the nu-Higgsinos}

\subsection{Production}

The nu-Higgsinos cannot be produced from decays of an MSSM Higgs, thus the primary production mechanism is through a Drell-Yan process.    The charged nu-Higgsino pair is produced through a Drell-Yan process involving a $\gamma$ or $Z$ boson while the neutral nu-Higgsino pair can only be produced through a $Z$.    Note that the $Z$ will only couple to two different nu-Higgsinos, but since the two neutral nu-Higgsinos are degenerate this will effectively be a pair production mechanism.    Associated production of a charged and neutral nu-Higgsino can occur through a Drell-Yan process involving the $W$ bosons.

\begin{figure}[htpb]
\begin{center}
\subfigure[]{\includegraphics[angle=0,width=0.5\textwidth]{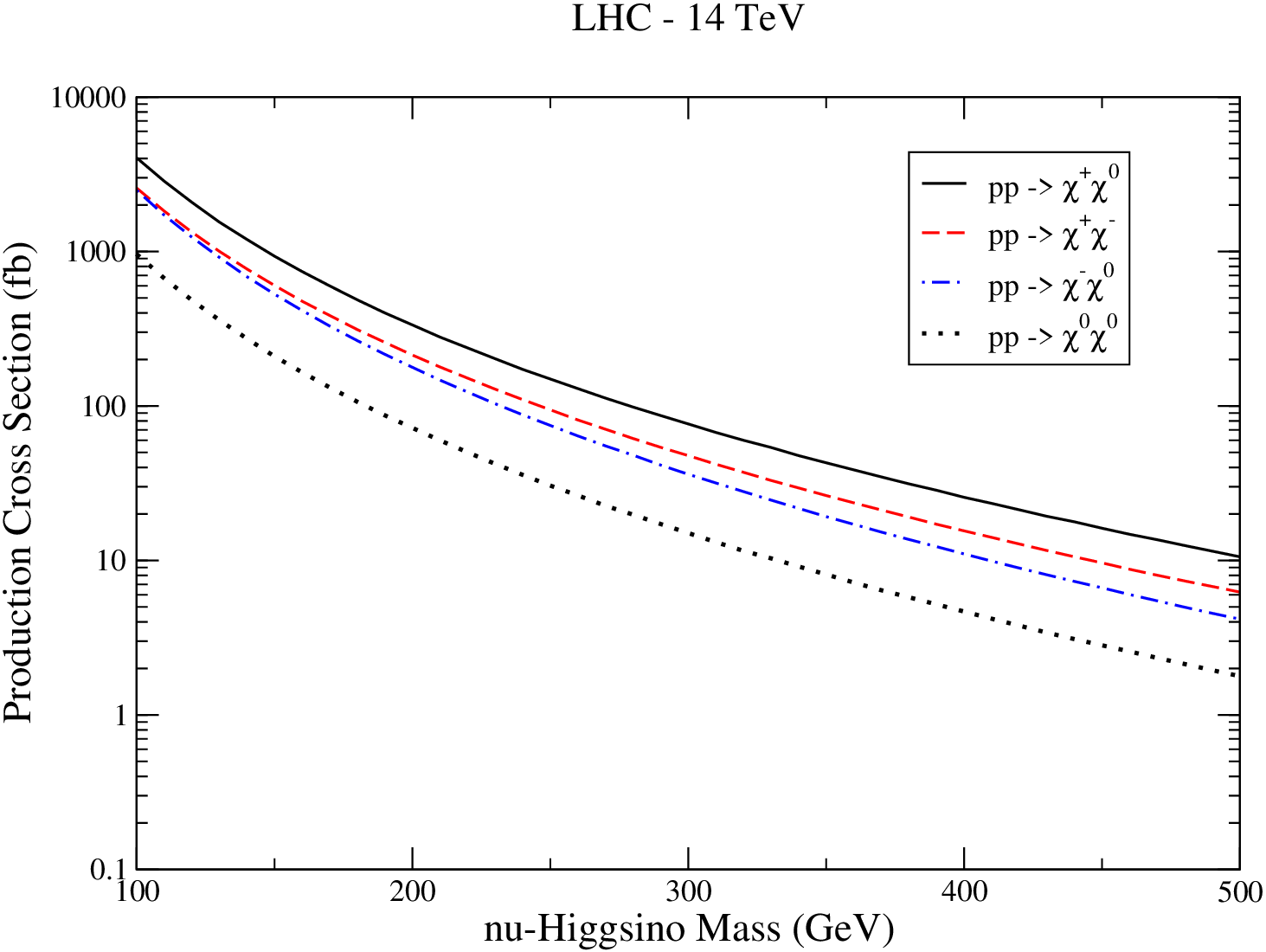}} \\
\vspace*{.13cm}
\subfigure[]{\includegraphics[angle=0,width=0.5\textwidth]{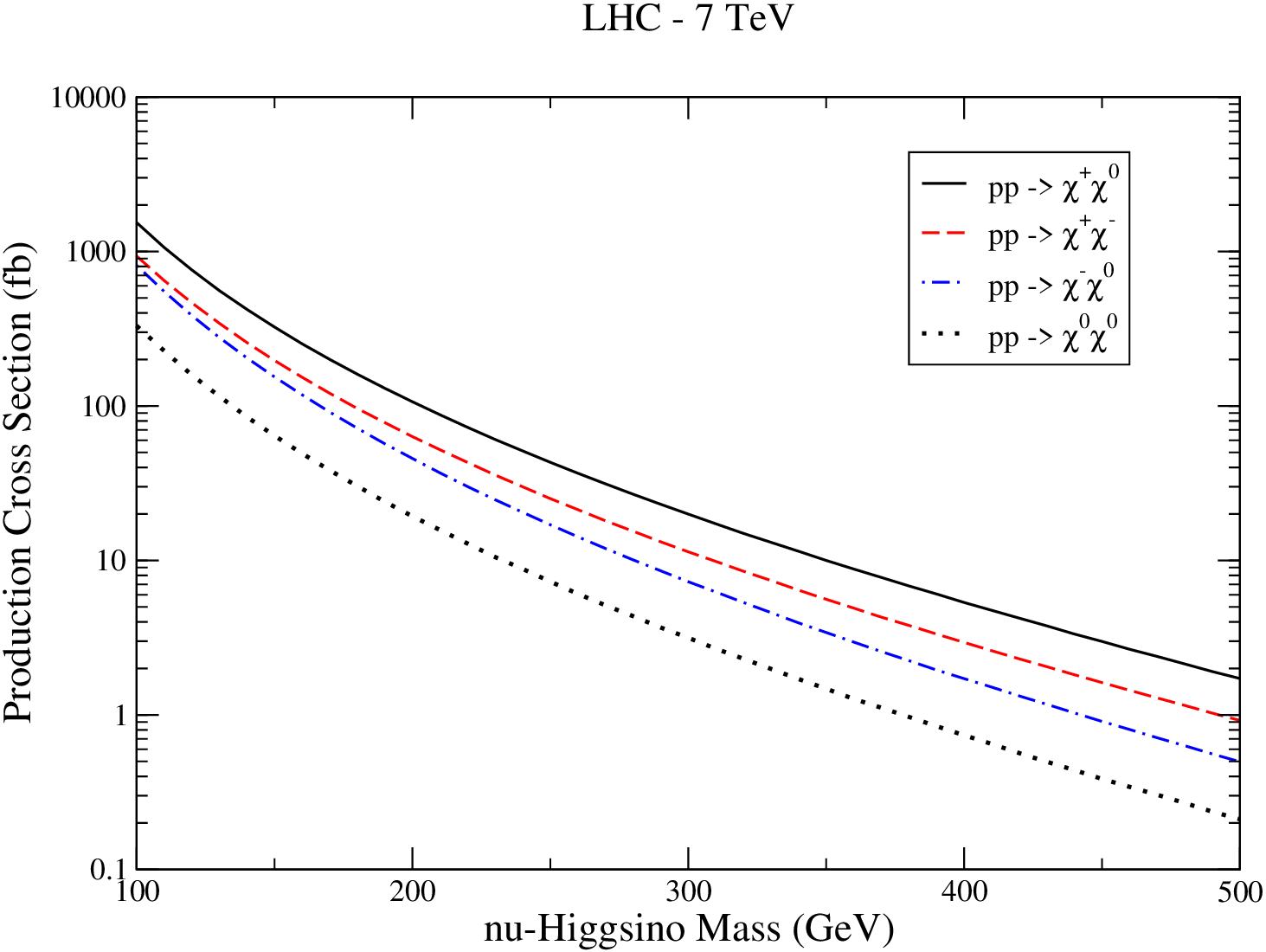}} \\
\vspace*{.13cm}
\subfigure[]{\includegraphics[angle=0,width=0.5\textwidth]{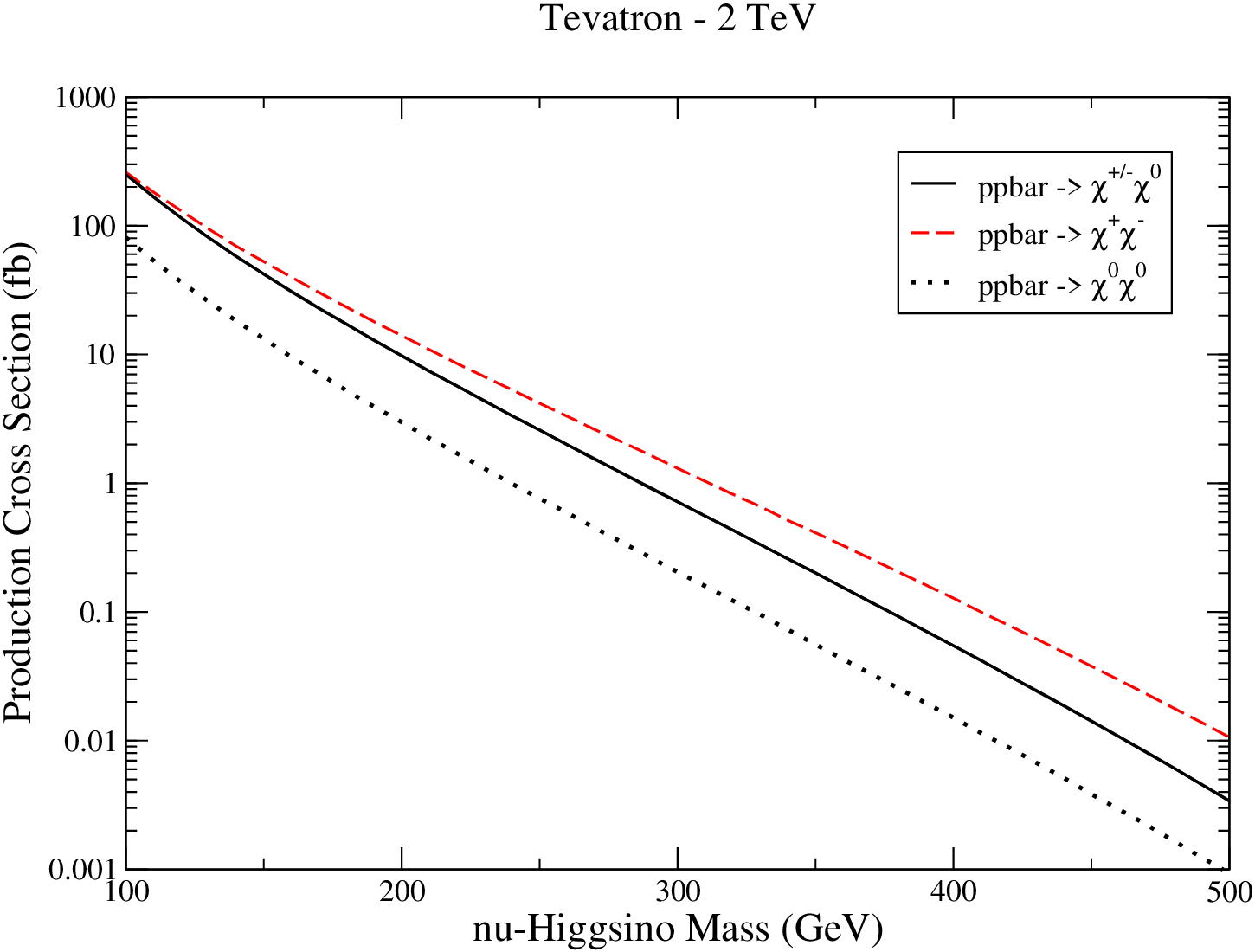}}
\caption{Production cross sections of various pairs of charged and neutral nu-Higgsinos (in fb) at the LHC with center of mass energies of (a) $14$ TeV and (b) $7$ TeV as well as the Tevatron with a center of mass energy of (c) $2$ TeV.}
\label{fig:higgsinoproduction}
\end{center}
\end{figure}

The LHC and Tevatron production cross sections of the above channels are calculated using the MadGraph/MadEvent software package~\cite{madgraph} with the CTEQ6L parton distribution functions~\cite{CTEQ}.    Using center of mass energies of $14$ and $7$ TeV for the LHC and $2$ TeV for the Tevatron we calculated the Higgsino pair production cross sections for Higgsino masses within the range of $100-500$ GeV.    The results are given in Fig~\ref{fig:higgsinoproduction}.    For the LHC, the $\widetilde{\chi}^{+}$-$\widetilde{\chi}^{0}$ pair has the largest production cross section.    Note that the $\widetilde{\chi}^{+}$-$\widetilde{\chi}^{0}$ cross section is larger than the $\widetilde{\chi}^{-}$/$\widetilde{\chi}^{0}$ production due to the valence up-quarks used in the former process as opposed to the valence down-quark in the latter.    We also note the larger charged nu-Higgsino pair production cross section compared to the neutral nu-Higgsino production from the Drell-Yan process involving a photon.

\subsection{Decays}

We first discuss the detailed decays of the neutral and then the charged nu-Higgsinos.     For simplicity, we will assume that the left and right handed sneutrinos have similar mass.   The $ \widetilde{\chi}^0$ decays into Branch A:  $ \nu_R +  \widetilde{\nu}_L$ or Branch B: $ \nu_L +  \widetilde{\nu}_R$ .      Since the mass of the neutrinos arises from the same Yukawa coupling term as this decay, the coupling is flavor diagonal and proportional to the mass.

Let us first consider Branch A.   The left-handed sneutrino decays are well-studied in the MSSM.   Here, we must choose a point in the mSUGRA parameter space.  We will choose Snowmass point SPS-1a, which is one of the most studied to date.  Using the results of Freitas et al~\cite{freitas},  we find that, for the normal hierarchy, the left-handed sneutrino decays into $ \widetilde{B}\nu_L \ 88\%$ of the time, either  $\nu_L \tau^+ \tau^-$ or $ \widetilde{B} \nu_L \tau^+\tau^- \ 7\%$ of the time, $\nu_L \tau^{\pm} \mu^{\mp} \ 5\%$ of the time, and has a negligible (O($0.2\%$)) decay into muon or electron pairs. For the inverted hierarchy, the left-handed sneutrino decays into $ \widetilde{B}\nu_L \ 88\%$ of the time, either $\nu_L \tau^+ \tau^-$ or $ \widetilde{B} \nu_L \tau^+\tau^- \ 5\%$ of the time, $\nu_L \tau^{\pm} e^{\mp} \ 5\%$ of the time, $\nu_L \tau^{\pm} \mu^{\mp} \ 2\%$ of the time, and also has a negligible (O($0.2\%$)) decay into muon or electron pairs. Since only $12\%$ of the decays are visible in either hierarchy, the branch into the left-handed sneutrino is not as promising as for the other branch.

Let us now consider Branch B:  The right-handed sneutrino decays have not been previously studied.  Since there are no gauge interactions, and negligible mixing with MSSM fields, the only decays of the right-handed sneutrino are three-body decays via a virtual nu-Higgsino.   One can have Branch B1: $\nu_L  \widetilde{\chi}^{0*}$ or Branch B2: $l  \widetilde{\chi}^{+*}$, where the $*$ superscript indicates a virtual state.  Branch B1 will give exactly the same flavor structure as Branch A, and thus we use those results.  For Branch B2, however, the final state will {\bf always} be visible, leading to the most interesting signatures.

Consider the normal hierarchy.  The right handed sneutrino has a specific flavor, and thus the flavor of the lepton in the decay leads to a muon $50\%$ of the time and a tau $50\%$ of the time (see Appendix A for a discussion).     As before, the $ \widetilde{\chi}^{+*}$ will decay with the same 50-50 split.    Thus, one expects Branch B2 to give $\mu \widetilde{\mu}$, $\mu \widetilde{\tau}$,$\tau \widetilde{\mu}$ and $\tau \widetilde{\tau}$ each $25\%$ of the time.  For the inverted hierarchy, the flavor of the lepton in the decay of the right handed electron sneutrino is always that of an electron, and the decay of the right handed muon or tau sneutrino is into  a muon $50\%$ of the time, and tau $50\%$ of the time. The $ \widetilde{\chi}^{+*}$ also decays with the 50-25-25 split in flavor so that branch B2 in the inverse hierarchy yields $e \widetilde{e}$, $e \widetilde{\mu}$, $e \widetilde{\tau}$, $\mu \widetilde{e}$, $\mu \widetilde{\mu}$, $\mu \widetilde{\tau}$, $\tau \widetilde{e}$, $\tau \widetilde{\mu}$, and $\tau \widetilde{\tau}$ with respective probabilities of $25\%$, $12.5\%$, $12.5\%$, $12.5\%$, $6.25\%$, $6.25\%$, $12.5\%$, $6.25\%$, and $6.25\%$. Finally, the slepton decays are given in Table 1 of Freitas et al.\cite{freitas}.

Putting all of these together, we find the decays of the $\widetilde{\chi}^0$  given in Table~\ref{tab:decayhiggsino-normal} and Table~\ref{tab:decayhiggsino-inverted}.  We have only included decays whose branching ratios exceed $0.5\%$.     Of course, each of the $\tau$ leptons will decay $17\%$ of the time into electrons and $17\%$ of the time into muons.

\begin{table}[ht]
\begin{center}
\begin{tabular}{|c|c|c|c|}
\hline
\multicolumn{2}{|c|}{$\widetilde{\chi}^{0}$} & \multicolumn{2}{c|}{$\widetilde{\chi}^{\pm}$} \\
\hline\hline Signature & Branching Fraction & Signature & Branching
Fraction \\ [0.5ex] \hline
Invisible &  66\% & $(\tau^{\pm})$ & 40\% \\
$\tau^{\pm} \mu^{\mp}$ & 14\% & $(\mu^{\pm})$ & 23\% \\
$\tau^{\pm} \tau^{\mp}$ & 12\% & $(\mu^{\pm}) \tau^{+} \tau^{-}$ & 12\% \\
$\mu^{\pm} \mu^{\mp}$ & 3.1\% & $(\tau^{\pm}) \tau^{+} \tau^{-}$ & 8.2\% \\
$\tau^{\pm} \tau^{+} \tau^{-} \mu^{\mp}$ & 2.0\% & $(\tau^{\pm}) \mu^{+} \mu^{-}$ & 4.7\% \\
$\tau^{+} \tau^{-} \mu^{+} \mu^{-}$ & 1.1\% & $(\tau^{\mp}) (\mu^{\pm} \mu^{\pm})$ & 2.9\% \\
$\tau^{+} \tau^{-} \tau^{+} \tau^{-}$ & 0.94\% & $(\mu^{\pm}) (\tau^{\pm} \tau^{\pm})$ & 2.9\% \\
& & $(\mu^{\pm}) \mu^{+} \mu^{-}$ & 1.8\% \\
& & $(\tau^{\pm}) \tau^{+} \tau^{-} \mu^{+} \mu^{-}$ & 1.0\% \\
& & $(\mu^{\pm}) \tau^{+} \tau^{-} \tau^{+} \tau^{-}$ & 1.0\% \\
& & $(\mu^{\pm}) \mu^{+} \mu^{-} \tau^{+} \tau^{-}$ & 0.56\% \\[1ex]
\hline
\end{tabular}
\caption{Decay branching fractions for the nu-Higgsinos assuming a normal hierarchy.  For the charged nu-Higgsino, the lepton in parenthesis has the upper sign for $\widetilde{\chi}^+$ and the lower sign for $\widetilde{\chi}^-$.}
\label{tab:decayhiggsino-normal}
\end{center}
\end{table}

We now turn to the decays of the $ \widetilde{\chi}^+$.   Here our work is already done from the discussion of Branch B2 above.    The resulting branching ratios are given in Tables~\ref{tab:decayhiggsino-normal} and~\ref{tab:decayhiggsino-inverted}.   Note that one can have single lepton, tri-lepton or penta-lepton decays.

\begin{table}[ht]
\begin{center}
\begin{tabular}{|c|c|c|c|}
\hline
\multicolumn{2}{|c|}{$\widetilde{\chi}^{0}$} & \multicolumn{2}{c|}{$\widetilde{\chi}^{\pm}$} \\
\hline\hline Signature & Branching Fraction & Signature & Branching
Fraction \\ [0.5ex] \hline
Invisible &  66\% & ($\tau^{\pm}$) & 28\% \\
$\tau^{\pm}e^{\mp}$ & 11\% & $(e^{\pm})$ & 23\% \\
$\tau^{+} \tau^{-}$ & 6.7\% & $(\mu^{\pm})$ & 11\% \\
$\tau^{\pm} \mu^{\mp}$ & 5.1\% & $(e^{\pm}) \tau^{+} \tau^{-}$ & 5.9\% \\
$e^{+} e^{-}$ & 3.1\% & $(\tau^{\pm}) e^{+} e^{-}$ & 4.1\% \\
$\mu^{\pm} e^{\mp}$ & 3.0\% & $(\mu^{\pm}) \tau^{+} \tau^{-}$ & 3.6\% \\
$\tau^{+} \tau^{-} e^{+} e^{-}$ & 1.1\% & $(e^{\pm}) e^{+} e^{-}$ & 3.3\% \\
$\tau^{+} \tau^{-} \mu^{\pm} e^{\mp}$ & 1.1\% & $(\tau^{\mp})(e^{\pm} e^{\pm})$ & 3.2\% \\
$ \tau^{+} \tau^{-} e^{\pm} \tau^{\mp}$ & 1.0\% & $(\tau^{\pm}) \tau^{+} \tau^{-}$ & 2.9\% \\
$\mu^{+} \mu^{-}$ & .91\% & $(\mu^{\pm}) \tau^{\pm} e^{\mp}$ & 2.1\% \\
 & & $(\mu^{\pm}) e^{+} e^{-}$ & 1.7\% \\
 & & $(e^{\mp}) (\tau^{\pm} \tau^{\pm})$ & 1.1\% \\
 & & $(e^{\pm}) e^{+} e^{-} \tau^{+} \tau^{-}$ & 1.1\% \\
 & & $(\tau^{\pm}) \mu^{+} \mu^{-}$ & 0.84\% \\
 & & $(\mu^{\mp}) (e^{\pm} e^{\pm})$ & 0.75\% \\
 & & $(e^{\pm}) \mu^{+} \mu^{-}$ & 0.69\% \\
 & & $(\mu^{\pm}) e^{+} e^{-} \tau^{+} \tau^{-}$ & 0.52\% \\
 & & $(\tau^{\mp})(\mu^{\pm} \mu^{\pm})$ & 0.51\% \\
 & & $(\mu^{\mp}) (\tau^{\pm} \tau^{\pm})$ & 0.51\% \\
& & $(\tau^{\pm}) e^{+} e^{-} \tau^{+} \tau^{-}$ & 0.50\% \\[1ex]
\hline
\end{tabular}
\caption{Decay branching fractions of nu-Higgsinos assuming an inverted hierarchy. For the charged nu-Higgsino, the lepton in parenthesis has the upper sign for $\widetilde{\chi}^+$ and the lower sign for $\widetilde{\chi}^-$.}
\label{tab:decayhiggsino-inverted}
\end{center}
\end{table}

\subsection{Signatures}

The high multiplicity of charged leptons in the nu-Higgsino decays gives some remarkable signatures.    From pair production of the $ \widetilde{\chi}^+ \widetilde{\chi}^-$, one can have very dramatic hexalepton and tetralepton events.    Associated production of $ \widetilde{\chi}^{\pm}$ with $ \widetilde{\chi}^0$ leads to pentalepton events, and pair production of $ \widetilde{\chi}^0 \widetilde{\chi}^0$ leads to tetralepton events.   The production cross sections were given earlier.

\subsubsection{Hexalepton events}

Perhaps the most dramatic events are hexalepton events which arise from the decay of a charged nu-Higgsino pair.   From Tables~\ref{tab:decayhiggsino-normal} and~\ref{tab:decayhiggsino-inverted}, one can see that the $\widetilde{\chi}^\pm$ has three charged leptons in the decay $31-32\%$ of the time, for either hierarchy, and thus the pair will yield six leptons roughly $10\%$ of the time.   Note that the production rate at the $14$~TeV LHC for a light $\widetilde{\chi}^\pm$  pair can be as large as 2600 fb, leading to an enormous hexalepton rate of over 200 fb.    

One will get a distribution of lepton flavors depending on the hierarchy.   How robust are these results?  If the $\widetilde{\chi}^\pm$  is lighter than the slepton or sneutrino, the flavor structure will not change--it simply means that the slepton or sneutrino is virtual, thus these results won't change.  We did choose a specific point in the mSUGRA parameter space, and that will change the flavor distribution, and thus these precise percentages should be taken {\it cum grano salis}.    Nonetheless, one does expect a huge production rate for hexalepton events if the $\widetilde{\chi}^\pm$  is not too heavy.

In Table~\ref{tab:cross-hexa}, we have summarized the signatures as follows.  In the first row, for both hierarchies and $\widetilde{\chi}^\pm$  masses of $100,200,300$~GeV, we give the total production rate for hexalepton events.   The tau's in the decays will either decay leptonically or hadronically.   If they decay hadronically, roughly $40\%$ will be identified.  If they decay leptonically, they will be indistinguishable from a muon or an electron.   In the remaining rows, we have listed the possible signatures, depending on the number of hadronic tau-identifications, weighted by the $40\%$ factor.    Since the electrons and muons are detected with virtually $100\%$ efficiency, {\bf all} of the decays listed in Table~\ref{tab:cross-hexa} are detectable.

\begin{table}[ht]
\begin{center}
\begin{tabular}{|c||c|c|c||c|c|c|} \hline
\multicolumn{1}{|c}{} & \multicolumn{3}{c}{Normal Hierarchy} & \multicolumn{3}{c|}{Inverted Hierarchy} \\
\hline\hline $M_{\widetilde{\chi}^{\pm}}$ & 100 GeV & 200 GeV & 300 GeV & 100 GeV & 200 GeV & 300 GeV \\ [0.5ex]
\hline
 $6$ Leptons & 260 & 26 & 5.1 & 240 & 24 & 4.8 \\
\hline
$0  \tau \ 6 \ell$ & 11 & 1.1 & 0.21 & 15 & 1.5 & 0.31 \\
$1  \tau \ 5 \ell$ & 17 & 1.7 & 0.35 & 20 & 2.0 & 0.39 \\
$2  \tau \ 4 \ell$ & 12 & 1.2 & 0.25 & 10 & 1.0 & 0.20 \\
$3  \tau \ 3 \ell$ & 4.6 & 0.46 & 0.092 & 2.5 & 0.25 & 0.050 \\
$4  \tau \ 2 \ell$ & 0.98 & 0.098 & 0.020 & 0.35 & 0.035 & 0.0070 \\
$5  \tau \ 1  \ell$ & 0.11 & 0.011 & 0.0020 & 0.024 & 0.0020 & 0.000 \\
$6  \tau \ 0  \ell$ & 0.0050 & 0.0010 & 0.000 & 0.0010 & 0.000 & 0.000 \\[1ex]
\hline
\end{tabular}
\caption{Cross sections, in femtobarns, for hexalepton signatures for various values of the charged nu-Higgsino mass and for the two different neutrino mass hierarchies.  The first line gives the total event rate into six leptons.   For the other lines, we include the fact that leptonic $\tau$ decays lead to additional $\mu$'s and $e$'s and that hadronic $\tau$ decays are detected with a $40\%$ efficiency.  In the table $\ell$ refers to either muons or electrons, and $\tau$ refers to an identified hadronically decaying tau.  The quoted cross sections are assuming a center of mass energy of $14$ TeV at the LHC.  A center of mass energy of $7$ TeV reduces the cross section roughly by a factor of $3$.}
\label{tab:cross-hexa}
\end{center}
\end{table}

One sees that each of the various signatures can be well within reach of the LHC, and some can have production cross sections of many tens of femtobarns.    We assumed here that $\sqrt{s}=14$~TeV.  If it is $7$~TeV, then the above figures show that the production rates are lower by a factor of roughly 3.    At the Tevatron, one can scale the results as shown in the production cross section figures.   One might be able to extend the sensitivity by consider the explicit charges.   For example, one combination in the normal hierarchy case would be $\mu^+\mu^+\mu^+\mu^-\tau^-\tau^-$ which might have lower backgrounds compared with three charge pairs.   Given the number of combinations, a detailed analysis of these possibilities would be premature.    Note that the total rate of identified hexalepton events, for a $\widetilde{\chi}^\pm$ mass of $100$~GeV is approximately 50 femtobarns at the LHC for $\sqrt{s}= 14$~TeV and approximately $16$ femtobarns for $\sqrt{s}= 7$~TeV.   At the Tevatron,  the rate would be roughly 4 fb, leading to a couple of dozen events for the current integrated luminosity.   Note that the production rate at the Tevatron will drop more quickly with the $\widetilde{\chi}^\pm$ mass than the LHC.

Suppose hexalepton events are not detected?    It could simply be that the charged nu-Higgsinos are too heavy, and thus failure to detect their decays would effectively place a lower limit on their mass.     Could a $100$~GeV charged nu-Higgsino evade detection?    The choice of hierarchy and the choice of the particular mSUGRA point will affect the flavor distribution (which can thus affect the tau content of the events and thus the efficiency of detection), but those will not substantially affect the size of the signal.    If the sneutrinos are heavy, this will also not affect the decays, unless they are so heavy that charged nu-Higgsino decays into a nu-Higgs plus the LSP can dominate, leading to only two leptons in the decay.   Perhaps the simplest way to reduce the signal is to have the right-handed sneutrino be substantially heavier than the left-handed sneutrino, since the latter decays are much more likely to be invisible.   Nonetheless, for most of parameter space, the charged nu-Higgsino mass is the primary factor in the signal rate.    Note also that, independent of any of these factors,  failure to detect these events at the Tevatron would imply a lower bound on the necessary luminosity to detect them at the LHC, as one can read off from the production cross sections.

\subsubsection{Tetralepton and pentalepton events}

$\widetilde{\chi}^\pm$  pair production can also lead to tetra-lepton events, as well as $\widetilde{\chi}^0$ pair production.   In Table~\ref{tab:cross-tetra}, we have listed the total production rate, as well as the various signatures, for these events.   Finally, associated production from a $W$ will lead to pentalepton events, and these are also shown in Table~\ref{tab:cross-penta}.

The cross sections are larger than for hexalepton events, however backgrounds will also be larger.  For example, a tetralepton signature of $\mu^+\mu^-e^+e^-$ would have backgrounds from real or virtual $\gamma$'s and $Z$'s.     One would expect the pentalepton signature to have smaller backgrounds, and yet the cross sections are still much larger.  They are much more sensitive to the flavor distribution, as one can see in the table.   It may very well be that the pentalepton signatures will be the most sensitive.  The cross sections are somewhat lower than for tetraleptons, but one would expect the backgrounds to be substantially lower.  The hexaleptons are more dramatic, with even smaller backgrounds, but the cross sections are substantially higher for the pentalepton events.

How sensitive is the Tevatron?   Consider a nu-Higgsino mass of $200$ GeV.  The cross section for pentaleptons at the LHC for $\sqrt{s}=14$~TeV is approximately $50$ fb, as seen in Table~\ref{tab:cross-penta}.   Comparing cross sections between Figure~\ref{fig:higgsinoproduction}a and~\ref{fig:higgsinoproduction}c, one sees that the cross section at the Tevatron is roughly a factor of $50$ smaller, giving a cross section of $1$ fb.    Thus, with $8$ fb$^{-1}$ integrated luminosity, one would expect $8$ events, {\bf iff} all $\tau$'s could be identified.   Obviously, they can't, and Table~\ref{tab:cross-penta} shows that events with five non-tau's occur at a rate of roughly $20\%$ of the total.    Of course, we have not done a detailed analysis.  Some hadronic $\tau$'s would be identified,  and some of the leptons might not be identified (some would not have enough transverse momentum, for example).   Although we are not aware of any backgrounds that large, a full simulation should be carried out.  It does appear that a lower bound of approximately $200$ GeV on the nu-Higgsino masses could be obtained at the Tevatron.  Certainly, at a mass of $100$ GeV, the event rate is a factor of $10$-$20$ larger and detection should be straightforward.

\begin{table}[ht]
\begin{center}
\begin{tabular}{|c||c|c|c||c|c|c|} \hline
\multicolumn{1}{|c}{} & \multicolumn{3}{c}{Normal Hierarchy} & \multicolumn{3}{c|}{Inverted Hierarchy} \\
\hline\hline $M_{\widetilde{\chi}^{\pm}}$ & 100 GeV & 200 GeV & 300
GeV & 100 GeV & 200 GeV & 300 GeV \\ [0.5ex] \hline
4 Leptons & 580 & 57 & 11 & 570 & 56 & 11 \\
\hline
$0  \tau \ 4  \ell$ & 27 & 2.6 & 0.52 & 120 & 11 & 2.3 \\
$1  \tau \ 3  \ell$ & 43 & 4.2 & 0.83 & 140 & 14 & 2.7 \\
$2  \tau \ 2  \ell$ & 24 & 2.3 & 0.46 & 47 & 4.6 & 0.92 \\
$3  \tau \ 1  \ell$ & 5.5 & 0.53 & 0.11 & 7.2 & 0.71 & 0.14 \\
$4  \tau \ 0  \ell$ & 0.44 & 0.043 & 0.010 & 0.37 & 0.037 & 0.0070 \\[1ex]
\hline
\end{tabular}
\caption{Cross sections, in femtobarns, for tetralepton signatures for various values of the charged nu-Higgsino mass and for the two different neutrino mass hierarchies.  The first line gives the total event rate into four leptons.   For the other lines, we include the fact that leptonic $\tau$ decays lead to additional $\mu$'s and $e$'s and that hadronic $\tau$ decays are detected with a $40\%$ efficiency.  In the table $\ell$ refers to either muons or electrons, and $\tau$ refers to an identified hadronically decaying tau.   The quoted cross sections are assuming a center of mass energy of $14$ TeV at the LHC.  A center of mass energy of $7$ TeV reduces the cross section roughly by a factor of $3$.}
\label{tab:cross-tetra}
\end{center}
\end{table}

\begin{table}[ht]
\begin{center}
\begin{tabular}{|c||c|c|c||c|c|c|} \hline
\multicolumn{1}{|c}{} & \multicolumn{3}{c}{Normal Hierarchy} & \multicolumn{3}{c|}{Inverted Hierarchy} \\
\hline\hline $M_{\widetilde{\chi}^{\pm}}$ & 100 GeV & 200 GeV & 300
GeV & 100 GeV & 200 GeV & 300 GeV \\ [0.5ex]
\hline
5 Leptons & 770 & 64 & 13 & 600 & 50 & 10 \\
\hline
$0  \tau \ 5  \ell$ & 29 & 2.4 & 0.50 & 110 & 9.0 & 1.8 \\
$1  \tau \ 4  \ell$ & 49 & 4.1 & 0.84 & 100 & 8.3 & 1.7 \\
$2  \tau \ 3  \ell$ & 36 & 3.0 & 0.61 & 41 & 3.4 & 0.69 \\
$3  \tau \ 2  \ell$ & 12 & 0.97 & 0.20 & 8.9 & 0.74 & 0.15 \\
$4  \tau \ 1  \ell$ & 1.8 & 0.15 & 0.031 & 0.76 & 0.063 & 0.013 \\
$5  \tau \ 0  \ell$ & 0.11 & 0.0090 & 0.0020 & 0.015 & 0.0010 & 0.000 \\[1ex]
\hline
\end{tabular}
\caption{Cross sections, in femtobarns, for pentalepton signatures for various values of the charged nu-Higgsino mass and for the two different neutrino mass hierarchies.  The first line gives the total event rate into five leptons.   For the other lines, we include the fact that leptonic $\tau$ decays lead to additional $\mu$'s and $e$'s and that hadronic $\tau$ decays are detected with a $40\%$ efficiency.  In the table $\ell$ refers to either muons or electrons, and $\tau$ refers to an identified hadronically decaying tau.   The quoted cross sections are assuming a center of mass energy of $14$ TeV at the LHC.  A center of mass energy of $7$ TeV reduces the cross section roughly by a factor of $3$.}
\label{tab:cross-penta}
\end{center}
\end{table}

\section{Conclusions}

The discovery of nonzero neutrino masses has led to a new type of two Higgs doublet model in which a second Higgs doublet couples only to the lepton doublets and right-handed neutrinos, leading to Dirac neutrino masses.   We consider a supersymmetrized version of this model, which thus contains four Higgs doublets.   A global symmetry which is only very weakly broken prevents substantial mixing between the additional doublets, called nu-Higgs multiplets, and the MSSM Higgs sector.  

We study two aspects of the phenomenology of these new states.  For the scalar nu-Higgs fields, a new production mechanism leads to the possibility of a $5\sigma$ detection with $100$ fb${^{-1}}$ of integrated luminosity.    The second aspect contains extremely exciting possibilities.  For the nu-Higgsino states, we find remarkable phenomenological signatures.  The most dramatic of these signatures are hexalepton events, which contain six charged leptons and missing energy and  which are produced with a cross section of up to $250$ fb for a nu-Higgsino mass of $100$~GeV.  Many of these leptons are tau's, and when we fold in the $40\%$ detection efficiency, we find roughly $50, 5, 1$ fb cross sections for events in which all six leptons can be identified for nu-Higgsino masses of $100, 200, 300$~GeV respectively.  For $\sqrt{s}=7$~TeV, these numbers are lower by a factor of three, and for the Tevatron are lower by another factor of four.   Thus, this model gives the exciting possibility of substantial multi-lepton events which can be detected at the LHC and may, for lower nu-Higgsino masses, be detectable at the Tevatron.

\section*{Acknowledgments}   We are very grateful to Vernon Barger, John Conway, Linda Dolan, Pavel Fileviez Perez, Tao Han, Heather Logan, Gabe Shaughnessy, and Xerxes Tata for very helpful discussions.  This work is supported in part by the National Science Foundation, Grants PHY-0755262 and PHY-0503584, by the U.S. Department of Energy under grants No. DE-FG02-95ER40896, DE-FG02-05ER41361, and by the Wisconsin Alumni Research Foundation.

\appendix
\section{PMNS Mixing Matrix Parameters}
\label{appendix}

The decay rate of $\chi^{+} \rightarrow \ell^{+}_{i} \nu_{j}$ (as well as the rate for supersymmetric versions) is proportional to the square of the $ij$-element of the neutrino mass matrix in the flavor eigenstate basis. It, in turn, is related to the physical neutrino masses through the PMNS matrix  $U_{ij}$ by  $m^{D}_{\nu_{ij}} = U^{\dag}_{ik}m_{\nu_{kl}}U_{lj}$, where $m^{D}_{\nu_{ij}}$ is the diagonalized neutrino mass matrix. This gives
\begin{equation} 
\Gamma(\chi^{+} \rightarrow \ell^{+}_{i} \nu_{j}) \propto m^{2}_{\nu_{ij}} = \left( U_{ik}m^{D}_{\nu_{kl}}U^{\dag}_{lj} \right)^{2}.
\label{Mass Relation}
\end{equation}
The diagonalizing matrix and diagonalized mass matrix are given by (in the limit where $\theta_{13}=0$)
\begin{eqnarray}
U_{ij} &=& \left(
  \begin{array}{ccc}
    \cos(\theta_{12}) & \sin(\theta_{12}) & 0 \\
    -\sin(\theta_{12})\cos(\theta_{23}) & \cos(\theta_{12})\cos(\theta_{23}) & \sin(\theta_{23}) \\
    \sin(\theta_{12})\sin(\theta_{23}) & -\cos(\theta_{13})\sin(\theta_{23}) & \cos(\theta_{23}) \\
  \end{array}
\right), \\
\textrm{and} \hspace{.2 in} m^{D}_{\nu_{ij}} &=&
\left(
  \begin{array}{ccc}
    m_{\nu_{1}} & 0 & 0 \\
    0 & m_{\nu_{2}} & 0 \\
    0 & 0 & m_{\nu_{3}} \\
  \end{array}
\right) .
\end{eqnarray}

Although there is a range of allowed values for $\theta_{12},~\theta_{13},~\text{and}~\theta_{23}$, we will choose them to be given by $34^{\circ},~0^{\circ},~\text{and}~45^{\circ}$, respectively.   We will also assume that the lightest neutrino has negligible mass (thus not considering the fully degenerate case).   The reason for these assumption is that, in determining branching ratios, we are already choosing a specific point in mSUGRA parameter space, and thus our quantitative results should not be taken too precisely.    Choosing the central values of the neutrino parameters keeps the presentation simple.     With these assumptions, the diagonalizing
matrix becomes
\begin{equation}
U_{ij} \approx \left(
  \begin{array}{ccc}
    .83 & .56 & 0 \\
    -.4 & .59 & .71 \\
    .4 & -.59 & .71 \\
  \end{array}
\right) .
\end{equation}
For the normal hierarchy we have $m_{\nu_{1}} \approx m_{\nu_{2}} \ll m_{\nu_{3}}\equiv M$, while for the inverted hierarchy we have $m_{\nu_{3}} \ll m_{\nu_{1}} \approx m_{\nu_{2}}\equiv M$.

We then find
\begin{equation}
\Gamma(\chi^{+} \rightarrow \ell^{+}_{i} \nu_{j}) \propto m^{2}_{ij} = \left\{
  \begin{array}{lr}
    M^{2}|U_{i3}U^{*}_{j3}|^{2} & \ \textrm{Normal Hierarchy}\\
    M^{2}|U_{i1}U^{*}_{j1} + U_{i2}U^{*}_{j2}  |^{2} & \ \textrm{Inverted
    Hierarchy}
  \end{array}
\right.
\end{equation}
The branching ratio BR$(\chi^{+} \rightarrow \ell^{+}_{i} \nu_{j}) = \Gamma(\chi^{+} \rightarrow \ell^{+}_{i} \nu_{j}) / \Gamma(\chi^{+} \rightarrow \textrm{any} \ \ell^{+} \nu)$ is given by $m^{2}_{ij} \ / \sum_{kl} m^{2}_{kl}$, which is easily evaluated as 
\begin{equation}
\textrm{BR}(\chi^{+} \rightarrow \ell^{+}_{i}
\nu_{j}) = \left(
  \begin{array}{ccc}
    0 & 0 & 0 \\
    0 & 25 \% & 25 \% \\
    0 & 25 \% & 25 \% \\
  \end{array}
\right) \hspace{.2 in} \textrm{for the normal hierarchy},
\end{equation}
\begin{equation}
\textrm{BR}(\chi^{+} \rightarrow \ell^{+}_{i} \nu_{j}) = \left(
  \begin{array}{ccc}
    50 \% & 0 & 0 \\
    0 & 12.5 \% & 12.5 \% \\
    0 & 12.5 \% & 12.5 \% \\
  \end{array}
\right) \hspace{.2 in} \textrm{for the inverted hierarchy}.
\end{equation}
In the case of final state neutrinos, one sums over all flavors
since they are not detected. All of the above holds for the
analogous decays involving SUSY partners: $\widetilde{\chi}^{+}
\rightarrow \widetilde{\ell}^{+}_{i} \nu_{j}$ and also will give us the flavor structure when, for example, a $\widetilde{\nu}_R$ of flavor $j$ decays into a $\widetilde{\chi}^{+*}$ and a lepton $\ell^-_i$

\renewcommand{\baselinestretch}{1.4}

\end{document}